\documentclass[12pt]{article}
\usepackage[dvips]{graphicx}
\usepackage{amsmath}
\usepackage{graphicx}
\usepackage{amsfonts}
\usepackage{amssymb}
\usepackage{epsfig}
\usepackage{subfigure}

\newcommand{\be}{\begin{equation}}
\newcommand{\ee}{\end{equation}}
\newcommand{\bea}{\begin{eqnarray}}
\newcommand{\eea}{\end{eqnarray}}
\newcommand{\ba}{\begin{array}}
\newcommand{\ea}{\end{array}}
\newcommand{\beas}{\begin{eqnarray*}}
\newcommand{\eeas}{\end{eqnarray*}}
\newcommand{\bes}{\begin{equation*}}
\newcommand{\ees}{\end{equation*}}

\newcommand{\nn}{\nonumber}

\def\i2           {\mbox{$\frac{i}{2}$}}

\begin{document}
\title{\bf Constraining Transmission And Reflection Probabilities By Using The Miller-Good Transformation And Its Applications}

\author{P. Boonserm$^{a}$\hspace{1mm} and M. Visser$^{b}$\\
$^{a}$ {\small {\em Department of Mathematics and Computer Science, Chulalongkorn University}}\\
{\small {\em Bangkok 10330, Thailand}}\\
$^{b}$ {\small {\em School of Mathematics, Statistics, and Operations Research, Victoria University of Wellington,}}\\
{\small {\em Wellington, New Zealand}}
}

\maketitle

\begin{abstract}
Transmission through and reflection from a potential barrier, and the very closely related issue of particle production from a parametric resonance, are topics of considerable general interest in quantum physics. We have developed a rather general bound on quantum transmission probabilities, and recently applied it to bounding the greybody factors of a Schwarzschild black hole. In this current paper, we take a different tack - we report a way of using the Miller-Good transformation (which maps an initial   equation to a final   equation for a different potential) to significantly generalize the previous bound. We then apply this general formalism in a very specific manner to derive a rigorous bound that is ``as close as possible" to the usual WKB estimate for barrier penetration.
\vspace{5mm}

\noindent{\bf Keywords: transmission, reflection, Bogoliubov coefficients, analytic bound}

\end{abstract}

\section{Introduction}
Following the discussion presented in \cite{mil}, consider the Schr\"{o}dinger equation,
\begin{equation}
u(x)'' + k(x)^{2}u(x) = 0,\label{se}
\end{equation}
where, $k(x)^{2} = 2m[E - V(x)]/\hbar^{2}$. As long as $V(x)$ tends to finite (possibly different) constants $V_{\pm\infty}$ at left and right spatial infinity, then for $E > \max\{V_{+\infty}, V_{-\infty}\}$ one can set up a one-dimensional scattering problem in a completely straightforward manner - see for example any of the standard references \cite{Dicke, Merzbacher, Landau, Shankar, Capri, Bohm, Messiah, Branson, Liboff}. The scattering problem is completely characterized by the transmission and reflection amplitudes ($t$ and $r$), though the most important aspects of the physics can be extracted from the transmission and reflection probabilities ($T = |t|^{2}$ and $R = |r|^{2}$). Relatively little work has gone into providing general analytic bounds on the transmission probabilities, (as opposed to approximate estimates), and prior to the work leading to the current article, the only known result as far as we have been able to determine is this \cite{1D, Bogoliubov}:\\

\noindent \textbf{Theorem 1.} \textit{Consider the Schr\"{o}dinger equation (\ref{se}). Let $h(x) > 0$ be some positive but otherwise arbitrary once - differentiable function. Then the transmission probability is bounded from below by}
\begin{equation}
T \geq \text{sech}^{2}\left\{\int_{-\infty}^{\infty}\frac{\sqrt{(h')^{2} + \left(k^{2} - h^{2}\right)^{2}}}{2h}dx\right\}.\label{gb}
\end{equation}
To obtain useful information one should choose asymptotic conditions on the function $h(x)$ so that the integral converges - otherwise one obtains the true but trivial result $T \geq \text{sech}^{2}\infty = 0$. (There is of course a related bound in the reflection probability, $R$, and if one works with the formally equivalent problem of parametric oscillations, a related bound on the resulting Bogoliubov coefficients and particle production.)

This quite remarkable bound was first derived in \cite{1D}, with further discussion and an alternate proof being provided in \cite{Bogoliubov}. These bounds were originally used as a technical step when studying a specific model for sonoluminescence \cite{Sono}, and since then have also been used to place limits on particle production in analogue spacetimes \cite{analogue} and resonant cavities \cite{Dodonov}, to investigate qubit master equations \cite{Hall}, and to motivate further general investigations of one-dimensional scattering theory \cite{Integral}. Most recently, these bounds have also been applied to the greybody factors of a Schwarzschild black hole \cite{grey}.

A slightly weaker, but much more tractable, form of the bound can be obtained by applying the triangle inequality. For $h(x) > 0$:
\begin{equation}
T \geq \text{sech}^{2}\left\{\frac{1}{2}\int_{-\infty}^{\infty}\left[|\ln(h)'| + \frac{|k^{2} - h^{2}|}{h}\right]dx\right\}.\label{weak}
\end{equation}
Five important special cases are:

\noindent 1. If we take $h = k_{\infty}$, where $k_{\infty} = \lim_{x \rightarrow \pm\infty}k(x)$, then we have \cite{mil, 1D, Bogoliubov}
\begin{equation}
T \geq \text{sech}^{2}\left\{\frac{1}{2k_{\infty}}\int_{-\infty}^{\infty}|k_{\infty}^{2} - k^{2}|dx\right\}.\label{c1}
\end{equation}
2. If we define $k_{+\infty} = \lim_{x \rightarrow +\infty}k(x) \neq k_{-\infty} = \lim_{x \rightarrow -\infty}k(x)$, and take $h(x)$ to be any function that smoothly and monotonically interpolates between $k_{-\infty}$ and $k_{+\infty}$, then we have
\begin{equation}
T \geq \text{sech}^{2}\left\{\frac{1}{2}\left|\ln\left(\frac{k_{+\infty}}{k_{-\infty}}\right)\right| + \frac{1}{2}\int_{-\infty}^{\infty}\frac{|k^{2} - h^{2}|}{h}dx\right\}.
\end{equation}
This is already more general than the most closely related result presented in \cite{1D, Bogoliubov}.

\noindent 3. If we have a single extremum in $h(x)$ then
\begin{equation}
T \geq \text{sech}^{2}\left\{\frac{1}{2}\left|\ln\left(\frac{k_{+\infty}k_{-\infty}}{h_{ext}^{2}}\right)\right| + \frac{1}{2}\int_{-\infty}^{\infty}\frac{|k^{2} - h^{2}|}{h}dx\right\}.
\end{equation}
This is already more general than the most closely related result presented in \cite{1D, Bogoliubov}.

\noindent 4. If we have a single minimum in $k^{2}(x)$, and choose $h^{2} = \max\{k^{2}, \Delta^{2}\}$, assuming $k_{\text{min}}^{2} \leq \Delta^{2} \leq k_{\pm\infty}^{2}$, (but still permitting $k_{\text{min}}^{2} < 0$, so we are allowing for the possibility of a classically forbidden region), then
\begin{equation}
T \geq \text{sech}^{2}\left\{\frac{1}{2}\ln\left(\frac{k_{+\infty}k_{-\infty}}{\Delta^{2}}\right) + \frac{1}{2\Delta}\int_{\Delta^{2} > k^{2}}|\Delta^{2} - k^{2}|dx\right\}.\label{c4}
\end{equation}
This is already more general than the most closely related result presented in \cite{1D, Bogoliubov}.

We emphasize the $\Delta$ is an adjustable parameter that can in principle be used to optimize the bound - furthermore it is this special case that has a close connection to the WKB-like bound we shall eventually derive later on in this article.

\noindent 5. If $k^{2}(x)$ has a single minimum and $0 < k_{\text{min}}^{2} < k_{\pm\infty}^{2}$, then
\begin{equation}
T \geq \text{sech}^{2}\left\{\frac{1}{2}\ln\left(\frac{k_{+\infty}k_{-\infty}}{k_{\text{min}}^{2}}\right)\right\}.\label{c5}
\end{equation}
This is the limit of (\ref{c4}) above as $\Delta \rightarrow k_{\text{min}} > 0$, and is one of the special cases considered in \cite{1D}.

\noindent Comment: In the current article we shall not be seeking to directly apply the general bound (\ref{gb}), its weakened form (\ref{weak}), or any of its specializations as given in (\ref{c1})-(\ref{c5}) above. Instead we shall be seeking to extend and generalize the bound to make it more powerful.

The tool we shall use to do this is the Miller-Good transformation \cite{mil, Good}.

\section{The Miller-Good transformation}
Consider the Schr\"{o}dinger equation (\ref{se}), and consider the substitution \cite{mil, Good}
\begin{equation}
u(x) = \frac{1}{\sqrt{X'(x)}}U(X(x)).\label{miller}
\end{equation}
We will want $X$ to be our ``new" position variable, so $X(x)$ has to be an invertible function, which implies (via, for instance, the inverse function theorem) that we need $dX/dx \neq 0$. In fact, since it is convenient to arrange things so that the variables $X$ and $x$ both agree as to which direction is left or right, we can without loss of generality assert $dX/dx > 0$, whence also $dx/dX > 0$.

\noindent Now compute (using the notation $U_{X} = dU/dX$):
\begin{equation}
u'(x) = U_{X}(X)\sqrt{X'} - \frac{1}{2}\frac{X''}{(X')^{3/2}}U(X),
\end{equation}
and
\begin{equation}
u''(x) = U_{XX}(X)(X')^{3/2} - \frac{1}{2}\frac{X'''}{(X')^{3/2}}U + \frac{3}{4}\frac{(X'')^{2}}{(X')^{5/2}}U.
\end{equation}
Insert this into the original Schr\"{o}dinger equation, $u(x)'' + k(x)^{2}u(x) = 0$, to see that
\begin{equation}
U_{XX} + \left\{\frac{k^{2}}{(X')^{2}} - \frac{1}{2}\frac{X'''}{(X')^{3}} + \frac{3}{4}\frac{(X'')^{2}}{(X')^{4}}\right\}U = 0,
\end{equation}
which we can write as
\begin{equation}
U_{XX} + K^{2}U = 0,\label{tse}
\end{equation}
with
\begin{equation}
K^{2} = \frac{1}{(X')^{2}}\left\{k^{2} - \frac{1}{2}\frac{X'''}{X'} + \frac{3}{4}\frac{(X'')^{2}}{(X')^{2}}\right\}.
\end{equation}
That is, a Schr\"{o}dinger equation in terms of $u(x)$ and $k(x)$ has been transformed into a completely equivalent Schr\"{o}dinger equation in terms of $U(X)$ and $K(X)$. You can also rewrite this as
\begin{equation}
K^{2} = \frac{1}{(X')^{2}}\left\{k^{2} + \sqrt{X'}\left(\frac{1}{\sqrt{X'}}\right)''\right\}.
\end{equation}
The particular combination
\begin{equation}
\sqrt{X'}\left(\frac{1}{\sqrt{X'}}\right)'' = -\frac{1}{2}\frac{X'''}{X'} + \frac{3}{4}\frac{(X'')^{2}}{(X')^{2}}\label{combi}
\end{equation}
shows up in numerous a priori unrelated branches of physics and is sometimes referred to as the ``Schwarzian derivative" \cite{Schwarzian}.\footnote{We should warn the reader that in about 3\% of cases ``Schwarzian'' is mis-spelled as ``Schwartzian". We have been guilty of this error ourselves in the past. Worse, this spelling error then sometimes propagates and leads to false ``folk history". The Schwarz in question is Karl Hermann Amadeus Schwarz (1843 - 1921), as in ``Cauchy–Schwarz inequality", with the name of this inequality being incorrectly spelled in some 25\% of cases.}

\begin{itemize}
\item As previously mentioned, to make sure the coordinate transformation $x \leftrightarrow X$ is well defined we must have $X'(x) > 0$. Let us call this $j(x) \equiv X'(x)$ with $j(x) > 0$. We can then write
\begin{equation}
K^{2} = \frac{1}{j^{2}}\left\{k^{2} - \frac{1}{2}\frac{j''}{j} + \frac{3}{4}\frac{(j')^{2}}{j^{2}}\right\}.
\end{equation}
Let us suppose that $\lim_{x \rightarrow \pm\infty}j(x) = j_{\pm\infty} \neq 0$, then $K_{\pm\infty} = k_{\pm\infty}/j_{\pm\infty}$, so if $k^{2}(x)$ has nice asymptotic behaviour allowing one to define a scattering problem, then so does $K^{2}(x)$.

\item Another possibly more useful substitution (based on what we saw with the Schwarzian derivative) is to set $J(x)^{-2} \equiv X'(x)$ with $J(x) > 0$. We can then write
\begin{equation}
K^{2} = J^{4}\left\{k^{2} + \frac{J''}{J}\right\}.
\end{equation}
Let us suppose that $\lim_{x \rightarrow \pm\infty}J(x) = J_{\pm\infty} \neq 0$, then $K_{\pm\infty} = k_{\pm\infty}J_{\pm\infty}^{2}$, so if $k^{2}(x)$ has nice asymptotic behaviour allowing one to define a scattering problem, so does $K^{2}(x)$.
\end{itemize}
These observations about the behaviour at spatial infinity lead immediately and naturally to the two key results:\\

\noindent\textbf{Theorem 2.} \textit{Suppose $j_{\pm\infty} = 1$, (equivalently, $J_{\pm\infty} = 1$). Then the ``potentials" $k^{2}(x)$ and $K^{2}(X)$ have the same reflection and transmission amplitudes, and same reflection and transmission probabilities.}\\

\noindent\textit{Proof:} This is automatic since $K_{\pm\infty} = k_{\pm\infty}$, so (\ref{se}) and the transformed equation (\ref{tse}) both have the same asymptotic plane-wave solutions. Furthermore the Miller-Good transformation (\ref{miller}) maps any linear combination of solutions of (\ref{se}) into the same linear combination of solutions of the transformed equation (\ref{tse}). QED.\\

\noindent\textbf{Theorem 3.} \textit{Suppose $j_{\pm\infty} \neq 1$, (equivalently, $J_{\pm\infty} \neq 1$). What is the relation between the reflection and transmission amplitudes, and reflection and transmission probabilities of the two ``potentials" $k^{2}(x)$ and $K^{2}(X)$? This is also trivial - the ``potentials" $k^{2}(x)$ and $K^{2}(X)$ have the same reflection and transmission amplitudes, and same reflection and transmission probabilities.}\\

\noindent\textit{Proof:} The only thing that now changes is that the properly normalized asymptotic states are distinct
\begin{equation}
\frac{\exp(ik_{\infty}x)}{\sqrt{k_{\infty}}} \leftrightarrow \frac{\exp(iK_{\infty}x)}{\sqrt{K_{\infty}}},
\end{equation}
but map into each other under the Miller-Good transformation. QED.

\section{Improved general bounds}
We already know (from theorem 1) that:
\begin{equation}
T \geq \text{sech}^{2}\left\{\int_{-\infty}^{\infty}\vartheta dx\right\}.
\end{equation}
Here $T$ is the transmission probability, and $\vartheta$ is the function
\begin{equation}
\vartheta = \frac{\sqrt{(h')^{2} + \left(k^{2} - h^{2}\right)^{2}}}{2h},
\end{equation}
with $h(x) > 0$. But since the scattering problems defined by $k(x)$ and $K(X)$ have the same transmission probabilities, we also have
\begin{equation}
T \geq \text{sech}^{2}\left\{\int_{-\infty}^{\infty}\bar{\vartheta}dX\right\},
\end{equation}
with
\begin{equation}
dX = X'dx = jdx,
\end{equation}
and
\begin{equation}
\bar{\vartheta} = \frac{\sqrt{(h_{X})^{2} + \left(K^{2} - h^{2}\right)^{2}}}{2h}.
\end{equation}
We now unwrap the definitions to yield
\begin{eqnarray}
\bar{\vartheta} &=& \frac{1}{2h}\sqrt{\left(\frac{h'}{X}\right)^{2} + \left[\frac{1}{j^{2}}\left\{k^{2} - \frac{1}{2}\frac{j''}{j} + \frac{3}{4}\frac{(j')^{2}}{j^{2}}\right\} - h^{2}\right]}\\
                &=& \frac{1}{2hj}\sqrt{(h')^{2} + \left[\frac{1}{j}\left\{k^{2} - \frac{1}{2}\frac{j''}{j} + \frac{3}{4}\frac{(j')^{2}}{j^{2}}\right\} - jh^{2}\right]^{2}}.
\end{eqnarray}
This permits us to derive the first form of our improved bound.\\
\\
\textbf{Improved bound 1.} \textit{$\forall h(x) > 0, \forall j(x) > 0$ we have}
\begin{equation}
T \geq \text{sech}^{2}\left\{\int_{-\infty}^{\infty}\frac{1}{2h}\sqrt{(h')^{2} + \left[\frac{1}{j}\left\{k^{2} - \frac{1}{2}\frac{j''}{j} + \frac{3}{4}\frac{(j')^{2}}{j^{2}}\right\} - jh^{2}\right]^{2}}dx\right\}.\label{I1}
\end{equation}
Since this new bound contains two freely specifiable functions, it is definitely stronger than the result we started from (\ref{gb}). The result is perhaps a little more manageable if we work in terms of the function $J$ instead of $j$. We follow the previous logic but now set
\begin{equation}
dX = X'dx = J^{-2}dx,
\end{equation}
and
\begin{eqnarray}
\bar{\vartheta} &=& \frac{\sqrt{(h_{X})^{2} + \left(K^{2} - h^{2}\right)^{2}}}{2h}\nn\\
                &=& \frac{1}{2h}\sqrt{\left(\frac{h'}{X}\right)^{2} + \left[J^{4}\left\{k^{2} + \frac{J''}{J}\right\} - h^{2}\right]^{2}}.
\end{eqnarray}
This now leads to the second form of the improved bound.\\
\\
\textbf{Improved bound 2.} \textit{$\forall h(x) > 0, \forall J(x) > 0$ we have}
\begin{equation}
T \geq \text{sech}^{2}\left\{\int_{-\infty}^{\infty}\frac{1}{2h}\sqrt{(h')^{2} + \left[J^{2}\left\{k^{2} + \frac{J''}{J}\right\} - \frac{h^{2}}{J^{2}}\right]^{2}}dx\right\}.\label{I2}
\end{equation}
A useful further modification is to substitute $h = HJ^{2}$, leading to the third form of the improved bound.\\
\\
\textbf{Improved bound 3.} \textit{$\forall H(x) > 0, \forall J(x) > 0$ we have}
\begin{equation}
T \geq \text{sech}^{2}\left\{\int_{-\infty}^{\infty}\frac{1}{2H}\sqrt{\left[H' + 2H\frac{J'}{J}\right]^{2} + \left[k^{2} + \frac{J''}{J} - H^{2}\right]^{2}}dx\right\}.\label{I3}
\end{equation}
Since $J(x) > 0$ we can without any loss of generality write
\begin{equation}
J(x) = \exp\left[\int\chi(x)dx\right],
\end{equation}
where $\chi(x)$ is real but otherwise unconstrained. This permits is to write a fourth form of the improved bound.\\
\\
\textbf{Improved bound 4.} \textit{$\forall H(x) > 0, \forall \chi(x)$ we have}
\begin{equation}
T \geq \text{sech}^{2}\left\{\int_{-\infty}^{\infty}\frac{1}{2H}\sqrt{\left[H' + 2H\chi\right]^{2} + \left[k^{2} + \chi^{2} + \chi' - H^{2}\right]^{2}}dx\right\}.\label{I4}
\end{equation}
Equations (\ref{I1}), (\ref{I2}), (\ref{I3}), and (\ref{I4}), are completely equivalent versions of our new bound. This improved bound is definitely more general than the original bounds reported in \cite{1D, Bogoliubov}. Furthermore, using totally different techniques involving Shabat-Zakharov systems of ODEs, we have verified the correctness of these improved bounds by obtaining a completely independent proof in references \cite{Shabat, phd}.

\section{WKB-like bound}
To manipulate this general bound into a form that is ``as close as possible" to the standard WKB-result, let us first take version 4 of our improved bound, (\ref{I4}), and use the triangle inequality to write it in a slightly weakened form.\\
\\
\textbf{Improved bound 5.} \textit{$\forall H(x) > 0, \forall \chi(x)$ we have}
\begin{equation}
T \geq \text{sech}^{2}\left\{\int_{-\infty}^{\infty}\left[\left|\frac{H'}{2H} + \chi\right| + \frac{\left|k^{2} + \chi^{2} + \chi' - H^{2}\right|}{2H}\right]dx\right\}.\label{I5}
\end{equation}
We shall choose the functions $H(x)$ and $\chi(x)$ to optimize the bound and render it into WKB-like form. To do this, introduce a positive parameter $\Delta$ and, similarly to what was done in obtaining (\ref{c4}), choose $H^{2} = \max\{k^{2}, \Delta^{2}\}$. We have in mind here the situation where $k^{2}(x)$ has a single minimum, and that $k_{\text{min}}^{2} \leq \Delta^{2} \leq k_{\pm\infty}^{2}$, (but still permitting $k_{\text{min}}^{2} < 0$, so we are allowing for the possibility of a classically forbidden region). Then
\begin{eqnarray}
T &\geq& \text{sech}^{2}\left\{\int_{k^{2} > \Delta^{2}}\left[\left|\frac{k'}{2k} + \chi\right| + \frac{\left|\chi^{2} + \chi'\right|}{2H}\right]dx\right.\nn\\
  &&     \left. + \int_{k^{2} < \Delta^{2}}\left[\left|\chi\right| + \frac{\left|k^{2} + \chi^{2} + \chi'\right| - \Delta^{2}}{2\Delta}\right]dx\right\}.
\end{eqnarray}
Note that if $\chi \rightarrow 0$, then this reduces exactly to (\ref{c4}), so the above is a genuine generalization of previous results. Now define $\kappa^{2} = \max\{0, -k^{2}\}$, so that $\kappa = |k|$ in the classically forbidden region $k^{2} < 0$, while $\kappa = 0$ in the classically allowed region $k^{2} > 0$.

Furthermore, choose $\chi = \kappa$. Then one has split the integral into three separate regions:
\begin{eqnarray}
T &\geq& \text{sech}^{2}\left\{\int_{k^{2} > \Delta^{2}}\left|\frac{k'}{2k}\right|dx + \int_{0 < k^{2} < \Delta^{2}}\frac{\left|k^{2} - \Delta^{2}\right|}{2\Delta}dx\right.\nn\\
  &&     \left. + \int_{k^{2} < 0}\left[\kappa + \frac{\left|\kappa' - \Delta^{2}\right|}{2\Delta}\right]dx\right\}.
\end{eqnarray}
The first of these integrals can be explicitly carried out, while the last can again be bounded by the triangle inequality. This leads to
\begin{eqnarray}
T &\geq& \text{sech}^{2}\left\{\ln\left(\frac{k_{\infty}}{\Delta}\right) + \int_{0 < k^{2} < \Delta^{2}}\frac{\left|k^{2} - \Delta^{2}\right|}{2\Delta}dx\right.\nn\\
  &&     \left. + \int_{k^{2} < 0}\kappa dx + \int_{k^{2} < 0}\left[\frac{\left|\kappa' - \Delta^{2}\right|}{2\Delta}\right]dx\right\}.
\end{eqnarray}
The last integral appearing above can now be explicitly carried out. Collecting terms and rearranging we obtain our WKB-like bound.\\
\\
\textbf{Improved bound 6.} \textit{Let $k_{\infty} \geq \Delta0$, then for a single hump potential with the notation discussed above}
\begin{equation}
T \geq \text{sech}^{2}\left\{\int_{k^{2} < 0}\kappa dx + \ln\left(\frac{k_{\infty}}{\Delta}\right) + \frac{\kappa_{\text{max}}}{\Delta} + \frac{\Delta L}{2} + \int_{0 < k^{2} < \Delta^{2}}\frac{\left|k^{2} - \Delta^{2}\right|}{2\Delta}dx\right\}.\label{I6}
\end{equation}
Here $L$ denotes the width of the classically forbidden region, and $\kappa_{\text{max}}$ denotes the maximum height of the barrier. The first term in this bound involves the standard WKB integral $\int\kappa dx$ over the forbidden region, justifying the appellation ``WKB-like". Note that this bound is genuinely an improvement over all other published bounds we know of, the closest analogue to the current result appearing in our article \cite{mil}. That result corresponds to taking $\Delta \rightarrow k_{\infty}$ in which case the current bound (\ref{I6}) reduces to the less restrictive statement
\begin{equation}
T \geq \text{sech}^{2}\left\{\int_{k^{2} < 0}\kappa dx + \frac{\kappa_{\text{max}}}{k_{\infty}} + \frac{k_{\infty}L}{2} + \int_{k^{2} > 0}\frac{\left|k_{\infty}^{2} - k^{2}\right|}{2k_{\infty}}dx\right\}.\label{delty}
\end{equation}
Another key point is that this WKB-like bound (\ref{I6}) contains contributions from the classically allowed region. (As in general there must be, potentials with no classically forbidden region still generically have nontrivial scattering.) Compare this with the standard WKB estimate:
\begin{equation}
T_{\text{WKB}} \approx \text{sech}^{2}\left\{\int_{k^{2} < 0}\kappa dx + \ln 2\right\}.
\end{equation}
This or related forms of the WKB approximation for barrier penetration are derived, for instance, in many standard textbooks \cite{Dicke, Merzbacher, Landau, Shankar, Capri, Bohm, Messiah, Branson, Liboff}. Under the usual conditions applying to the WKB approximation for barrier penetration we have $\int\kappa dx \gg 1$, in which case one obtains the more well-known version
\begin{equation}
T_{\text{WKB}} \approx \exp\left\{-2\int_{k^{2} < 0}\kappa dx\right\}.
\end{equation}
The bound in (\ref{I6}), and it specialization in (\ref{delty}), are the closest we have so far been able to get to obtaining a rigorous bound that somewhat resembles the standard WKB estimate. Note the long chain of inequalities leading to these results - this suggests that the final inequality (\ref{I6}) might not be optimal. We are continuing to search for improvements on this WKB-like bound.

\section{Schwarzian bound}
In counterpoint to the WKB-like bound obtained above, there is also a rather elegant bound that can in certain circumstances be obtained in terms of the Schwarzian. First, take $h =$ (constant) in (\ref{I2}). Then
\begin{equation}
T \geq \text{sech}^{2}\left\{\frac{1}{2}\int_{-\infty}^{\infty}\left|\frac{J^{2}}{h}\left\{k^{2} + \frac{J''}{J}\right\} - \frac{h}{J^{2}}\right|dx\right\}.
\end{equation}
In order for this bound to convey nontrivial information we need $\lim_{x \rightarrow \pm\infty}J^{4}k^{2} = h^{2}$, otherwise the integral diverges and the bound trivializes to $T \geq \text{sech}^{2}(\infty) = 0$. The further specialization of this result reported in \cite{1D, Bogoliubov} and (\ref{c1}) above correspond to taking $J =$ (constant) $= \sqrt{h/k_{\infty}}$, which clearly is a weaker bound than that reported here. In the present situation we can without loss of generality set $h \rightarrow k_{\infty}$ in which case
\begin{equation}
T \geq \text{sech}^{2}\left\{\frac{1}{2}\int_{-\infty}^{\infty}\left|\frac{J^{2}}{k_{\infty}}\left\{k^{2} + \frac{J''}{J}\right\} - \frac{k_{\infty}}{J^{2}}\right|dx\right\}.\label{hty}
\end{equation}
We now need $\lim_{x \rightarrow \pm\infty}J = 1$ in order to make the integral converge. If $k^{2} > 0$, so that there is no classically forbidden region, then we can choose $J = \sqrt{k_{\infty}/k}$, in which case
\begin{equation}
T \geq \text{sech}^{2}\left\{\frac{1}{2}\int_{-\infty}^{\infty}\left|\frac{1}{\sqrt{k}}\left(\frac{1}{\sqrt{k}}\right)''\right|dx\right\}.\label{jkty}
\end{equation}
This is a particularly elegant bound in terms of the Schwarzian derivative, (\ref{combi}), which however unfortunately fails if there is a classically forbidden region. This bound is also computationally awkward to evaluate for specific potentials. Furthermore, in the current context there does not seem to be any efficient or especially edifying way of choosing $J(x)$ in the forbidden region, so while the bound in (\ref{hty}) is both elegant and explicit it is not particularly useful.

\section{Summary and discussion}
The fundamental bounds presented in this note are generally not ``WKB-like" - only for one particular special case has it proved useful to separate the region of integration into classically allowed and classically forbidden regions. In fact, in this WKB-like case it still seems that the bound we have derived is sub-optimal, and it is far from clear how closely these bounds might ultimately be related to WKB estimates of the transmission probabilities. This is an issue to which we hope to return in the future.\\
\\
We should mention that if one works with the formally equivalent problem of a parametric oscillator in the time domain then the relevant differential equation is
\begin{equation}
\ddot{u}(t) + k(t)^{2}u(t) = 0,
\end{equation}
and instead of asking questions about transmission amplitudes and probabilities one is naturally driven to ask formally equivalent questions about Bogoliubov coefficients and particle production. The key translation step is to realize that there is an equivalence \cite{mil, 1D, Bogoliubov, phd}:
\begin{equation}
T \leftrightarrow \frac{1}{1 + N}; ~~~ N \leftrightarrow \frac{1 - T}{T}.
\end{equation}
This leads to bounds on the number of particles produced that are of the form
\begin{equation}
N \leq \sinh^{2}\left\{\text{some appropriate integral}\right\}.
\end{equation}
To be more explicit about this, recall that our new improved bound can be written in any of four equivalent forms:\\
\\
1. For all $H(x) > 0$, for all $\chi(x)$ we have
\begin{equation}
T \geq \text{sech}^{2}\left\{\int_{-\infty}^{\infty}\frac{1}{2H}\sqrt{\left[H' + 2H\chi\right]^{2} + \left[k^{2} + \chi^{2} + \chi' - H^{2}\right]}dx\right\}.
\end{equation}
2. For all $H(x) > 0$, for all $J(x) > 0$,
\begin{equation}
T \geq \text{sech}^{2}\left\{\int_{-\infty}^{\infty}\frac{1}{2H}\sqrt{\left[H' + 2H\frac{J'}{J}\right]^{2} + \left[k^{2} + \frac{J''}{J} - H^{2}\right]^{2}}dx\right\}.
\end{equation}
3. For all $h(x) > 0$, for all $J(x) > 0$,
\begin{equation}
T \geq \text{sech}^{2}\left\{\int_{-\infty}^{\infty}\frac{1}{2h}\sqrt{(h')^{2} + \left[J^{2}\left\{k^{2} + \frac{J''}{J}\right\} - \frac{h^{2}}{J^{2}}\right]^{2}}dx\right\}.
\end{equation}
4. For all $h(x) > 0$, for all $j(x) > 0$,
\begin{equation}
T \geq \text{sech}^{2}\left\{\int_{-\infty}^{\infty}\frac{1}{2h}\sqrt{(h')^{2} + \left[\frac{1}{j}\left\{k^{2} - \frac{1}{2}\frac{j''}{j} + \frac{3}{4}\frac{(j')^{2}}{j^{2}}\right\} - jh^{2}\right]^{2}}dx\right\}.
\end{equation}
The equivalent statements about particle production are:\\
\\
1. For all $H(t) > 0$, for all $\chi(t)$ we have
\begin{equation}
N \leq \sinh^{2}\left\{\int_{-\infty}^{\infty}\frac{1}{2H}\sqrt{\left[H' + 2H\chi\right]^{2} + \left[k^{2} + \chi^{2} + \chi' - H^{2}\right]^{2}}dt\right\}.
\end{equation}
2. For all $H(t) > 0$, for all $J(t) > 0$,
\begin{equation}
N \leq \sinh^{2}\left\{\int_{-\infty}^{\infty}\frac{1}{2H}\sqrt{\left[H' + 2H\frac{J'}{J}\right]^{2} + \left[k^{2} + \frac{J''}{J} - H^{2}\right]^{2}}dt\right\}.\label{N2}
\end{equation}
3. For all $h(t) > 0$, for all $J(t) > 0$,
\begin{equation}
N \leq \sinh^{2}\left\{\int_{-\infty}^{\infty}\frac{1}{2h}\sqrt{(h')^{2} + \left[J^{2}\left\{k^{2} + \frac{J''}{J}\right\} - \frac{h^{2}}{J^{2}}\right]^{2}}dt\right\}.
\end{equation}
4. For all $h(t) > 0$, for all $j(t) > 0$,
\begin{equation}
N \leq \sinh^{2}\left\{\int_{-\infty}^{\infty}\frac{1}{2h}\sqrt{(h')^{2} + \left[\frac{1}{j}\left\{k^{2} - \frac{1}{2}\frac{j''}{j} + \frac{3}{4}\frac{(j')^{2}}{j^{2}}\right\} - jh^{2}\right]^{2}}dt\right\}.\label{N4}
\end{equation}
In closing, we reiterate that these general bounds reported in (\ref{I1}), (\ref{I2}), (\ref{I3}), and (\ref{I4}), their specializations in (\ref{I6}), (\ref{delty}), (\ref{hty}), and (\ref{jkty}), and the equivalent particle production bounds in (\ref{N2})-(\ref{N4}), are all general purpose tools that are applicable to a wide variety of physical situations \cite{Sono, analogue, Dodonov, Hall, Integral, grey}. Furthermore we strongly suspect that further generalizations of these bounds are still possible.

\section*{Acknowledgement}
This research was supported by the Marsden Fund, administered by the Royal Society of New Zealand. PB was additionally supported by a scholarship from the Royal Government of Thailand, and partially supported by a travel grant from FQXi, and by a grant for the professional development of new academic staff from the Ratchadapisek Somphot Fund at Chulalongkorn University, and by Thailand Toray Science Foundation (TTSF), and by the Research Strategic plan program (A1B1), Faculty of Science, Chulalongkorn University. The authors wish to thank the referee for several useful suggestions regarding the presentation and interpretation of the results.


\begin{thebibliography}{}
\bibitem{mil}
P. Boonserm and M. Visser, ``Transmission probabilities and the Miller-Good transformation," Journal of Physics A: Mathematical and Theoretical 42, 045301, 2009 [arXiv: 0806.2209 [gr-qc]].

\bibitem{Dicke}
R. H. Dicke and J. P. Wittke, Introduction to Quantum Mechanics. Addison Wesley, Reading: 1960.

\bibitem{Merzbacher}
E. Merzbacher, Quantum Mechanics. Wiley, New York: 1961.

\bibitem{Landau}
L. D. Landau and E. M. Lifschitz, Quantum Mechanics. Pergamon, Oxford: 1977.

\bibitem{Shankar}
R. Shankar, Principles of Quantum Mechanics. Plenum, New York: 1980.

\bibitem{Capri}
A. Z. Capri, Nonrelativistic Quantum Mechanics. Benjamin/Cummings, Menlo Park: 1985.

\bibitem{Bohm}
D. Bohm, Quantum Theory, Dover. Mineola: 1989.

\bibitem{Messiah}
A. Messiah, Quantum Mechanics. Dover, Mineola: 1999.

\bibitem{Branson}
B. H. Branson and C. J. Joachim, Quantum Mechanics. Prentice Hall/Pearson, Harlow: 2000.

\bibitem{Liboff}
R. L. Liboff, Introductory Quantum Mechanics. AddisonWesley, San Francisco: 2003.

\bibitem{1D}
M. Visser, ``Some general bounds for 1-D scattering," Phys. Rev. A 59, 427, 1999 [arXiv:quant-ph/9901030].

\bibitem{Bogoliubov}
P. Boonserm and M. Visser, ``Bounding the Bogoliubov coefficients," Annals Phys. 323, 2779, 2008 [arXiv:0801.0610 [quant-ph]].

\bibitem{Sono}
S. Liberati, M. Visser, F. Belgiorno and D. W. Sciama, ``Sonoluminescence as a QED vacuum effect. 1: The physical scenario," Phys. Rev. D 61, 085023, 2000 [arXiv:quant-ph/9904013].\\
S. Liberati, M. Visser, F. Belgiorno and D. W. Sciama, ``Sonoluminescence as a QED vacuum effect. 2: Finite volume effects," Phys. Rev. D 61, 085024, 2000 [arXiv:quantph/9905034].\\
S. Liberati, ``Quantum vacuum effects in gravitational fields: Theory and detectability," arXiv:gr-qc/0009050.

\bibitem{analogue}
C. Barcel´o, S. Liberati and M. Visser, ``Probing semiclassical analogue gravity in Bose-Einstein condensates with widely tunable interactions," Phys. Rev. A 68, 053613, 2003 [arXiv:cond-mat/0307491].\\
P. Jain, S. Weinfurtner, M. Visser and C. W. Gardiner, ``Analogue model of a FRW universe in Bose-Einstein condensates: Application of the classical field method," Phys. Rev. A 76, 033616, 2007 [arXiv:0705.2077 [cond-mat.other]].\\
S. Weinfurtner, ``Emergent spacetimes," arXiv:0711.4416 [gr-qc].\\
S. Weinfurtner, P. Jain, M. Visser and C. W. Gardiner, ``Cosmological particle production in emergent rainbow spacetimes," arXiv:0801.2673 [gr-qc].\\
S. Weinfurtner, A. White and M. Visser, ``Trans-Planckian physics and signature change events in Bose gas hydrodynamics," Phys. Rev. D 76, 124008, 2007 [arXiv:grqc/0703117].

\bibitem{Dodonov}
A. V. Dodonov, E. V. Dodonov, and V. V. Dodonov, ``Photon generation from vacuum in nondegenerate cavities with regular and random periodic displacements of boundaries," Physics Letters. A 317, 378-388, 2003 [arXiv: quant-ph/0308144v1].

\bibitem{Hall}
M. J. W. Hall, ``Complete positivity for time-dependent qubit master equations," J. Phys. A: Math. Theor. 41, 205302, 2008 [arXiv: 0802.0606v2 [quant-ph]].

\bibitem{Integral}
V. E. Barlette, M. M. Leite, and S. K. Adhikari, ``Integral equations of scattering in one dimension," Am. J. Phys. 69, 1010-1013, 2001 [arXiv:quant-ph/0103018].\\
L. L. S´anchez-Soto, J. F. Cari˜nena, A. G. Barriuso, and J. J. Monz´on, ``Vectorlike representation of one-dimensional scattering," Eur. J. Phys. 26, 469480, 2005 [arXiv:quant-ph/0411081].\\
T. R. Yang, M. M. Dvoynenko, A. V. Goncharenko, and V. Z. Lozovski, ``An exact solution of the Lippmann-Schwinger equation in one dimension," Am. J. Phys. 71, 64-71, 2003.

\bibitem{grey}
P. Boonserm and M. Visser, ``Bounding the greybody factors for Schwarzschild blackholes," Phys. Rev. D 78, 101502, 2008 arXiv:0806.2209 [gr-qc].

\bibitem{Good}
S. C. Miller and R. H. Good, ``A WKB-type approximation to the Schr\"{o}dinger equation," Phys. Rev. 91, 174-179, 1953.

\bibitem{Schwarzian}
V. Ovsienko and S. Tabachnikov, ``What is the Schwarzian derivative?," Notices of the AMS, 56, 34-35, 2009.

\bibitem{Shabat}
P. Boonserm and M. Visser, ``Reformulating the Schr\"{o}dinger equation as a Shabat-Zakharov system," arXiv:0910.2600 [math-ph].

\bibitem{phd}
P. Boonserm, ``Rigorous bounds on Transmission, Reflection, and Bogoliubov coefficients", PhD thesis (Victoria University of Wellington), arXiv:0907.0045 [math-ph].
\end{thebibliography}
\end{document}